\documentclass[referee]{cjaa}           

\usepackage{graphicx}                   
\input{epsf.sty}                        
\input{psfig.sty}                       

\begin{document}

   \title{Numerical Simulations Reveal the Origin of QPOs in Black Hole Candidates
}

   \volnopage{Vol.0 (200x) No.0, 000--000}      
   \setcounter{page}{1}          

   \author{Sandip K. Chakrabarti
      \inst{1,2}\mailto{}
      }

   \institute{S. N. Bose National Centre for Basic Sciences, Salt Lake, Kolkata, 700098, India\\
             \email{chakraba@bose.res.in}
\and
Centre for Space Physics, Chalantika 43, Garia Station Road, Kolkata 700084, India\\ }

   \date{Received~~2004 July 30; accepted~~2004~~month day}

   \abstract{We present results of various types of numerical simulations
of black hole accretion disks and find that those flows which are relatively non-dissipative and
which contain accretion shocks are the best candidates so far. The power density spectra (PDS)
reveal the aspects which are similar to what are observed in black hole candidates.
}
   \authorrunning{ S. K. Chakrabarti}            
   \titlerunning{Origin of QPOs}  

   \maketitle

\noindent To be Published in the proceedings of the 5th Microquasar Conference: 
Chinese Journal of Astronomy and Astrophysics


%
%
\section{Introduction}           
\label{sect:intro}

Quasi-Periodic Oscillations (QPOs) are the most puzzling aspects of black hole
astrophysics because they remain unexplained by popular models of accretion 
disks around a black hole. There is no doubt that the oscillation involves
the dynamical and non-linear variation of certain region of the accretion disk itself.
Thus, even though there has been popular euphoria about `models' which can `explain' the
`frequencies' of QPOs by vibrations of the disk, or Keplerian frequency at the
inner stable orbit, or trapped oscillations, etc. these models can be numerology 
at the best. By definition, any perturbation would oscillate with a frequency close 
or equal to the Keplerian frequency, whether it is perturbed vertically or horizontally 
or any other way. So, explaining the frequency is not a big problem. The major problem is: does
any solution (not a model) actually reproduce the power density spectra, break frequency, the 
power at the QPO frequency, even the multitude of QPOs observed on the same day, {\it and} 
explain the variation of QPO frequency with luminosity etc. In this short paper, 
we will show that if we stick to the proper solutions of the governing  equations, 
the result is very hopeful even when simple assumptions about flow configuration, 
cooling properties etc. are made. In future, when cooling processes are better handled,
we are convinced that 
we should be able to explain all the aspects of the QPOs very satisfactorily without taking resort to
particle dynamics or poorly defined models.

\section{Early Numerical Solutions}

The first successful numerical simulation was carried out in 1994 (Molteni, Sponholz and Chakrabarti, 1996,
hereafter MSC96). Theoretical works (Chakrabarti, 1989, 1990). Early numerical simulations
(Chakrabarti \& Molteni, 1993; Molteni, Lanzafame \& Chakrabarti, 1994) clearly showed that the centrifugal
barrier causes axisymmetric standing shocks to form around black holes and further theoretical 
(Chakrabarti, 1990) and numerical simulations (Lanzafame, Molteni \& Chakrabarti, 1997)
indicated that when the viscosity is high enough, the standing shock disappears. 
Figures 1(a-b) show two stages of the accretion flow configuration at half-cycle interval
in the simulation of MSC96. If we denote the  region between the
innermost sonic point and the shock as the CENtrifugal barrier dominated BOundary Layer or CENBOL, then 
the oscillation of the CENBOL takes place in place of cooling.  
Here, a power-law cooling $\propto \rho^2T^\alpha$ was used, $\alpha=0.5$ corresponds to
bremsstrahlung. Physically, cooling reduces the post-shock pressure and moves the shock
inward at a steady location. But when it is perturbed, say pushed backward, the post-shock temperature
goes up as the relative motion between the pre-shock and post-shock rises. This causes excess cooling
and the shock collapses toward the black hole only to be bounced back due to centrifugal force. 
The calculation of infall time becomes problematic since the presence of 
turbulence causes the flow to deviate from falling freely. On an average, it is seen that a constant velocity
in the post-shock region, at least up to the inner sonic point, could be a better guess 
(Chakrabarti \& Manickam, 2000).

\begin{figure}
   \begin{center}
\vskip -0.0cm
   \mbox{\epsfxsize=0.4\textwidth\epsfysize=0.3\textwidth\epsfbox{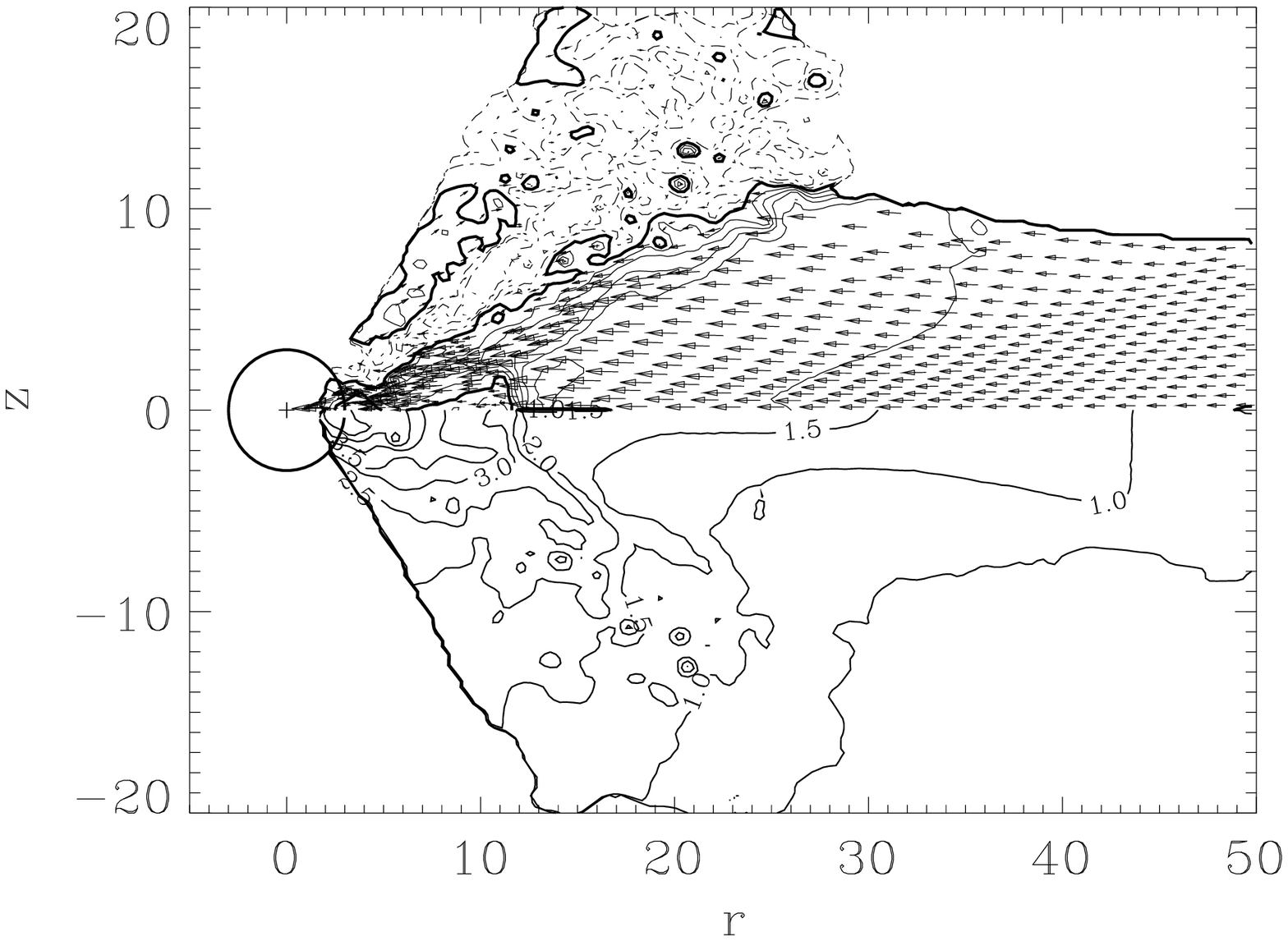}}
\vspace{0 cm}
   \mbox{\epsfxsize=0.4\textwidth\epsfysize=0.3\textwidth\epsfbox{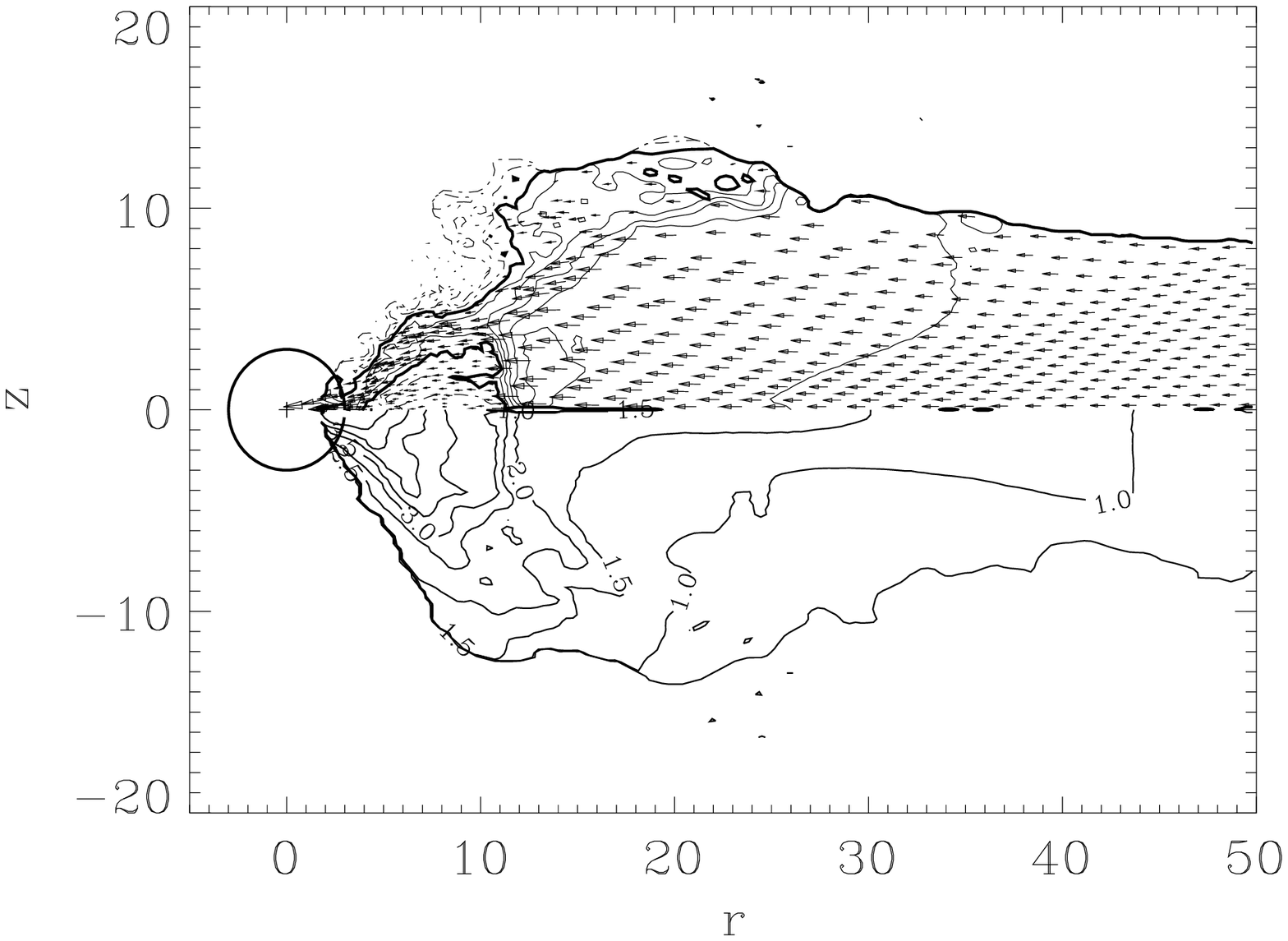}}
\vskip -0.0cm
   \caption{Comparison of the flow topologies at two phases of the shock oscillations
in simulations in which only the radial oscillation of the shocks are allowed. (from MSC96). 
Power law cooling has been used.} 
   \end{center}
\end{figure}

\begin{figure}
   \begin{center}
\vskip 0.0cm
   \mbox{\epsfxsize=0.8\textwidth\epsfysize=0.6\textwidth\epsfbox{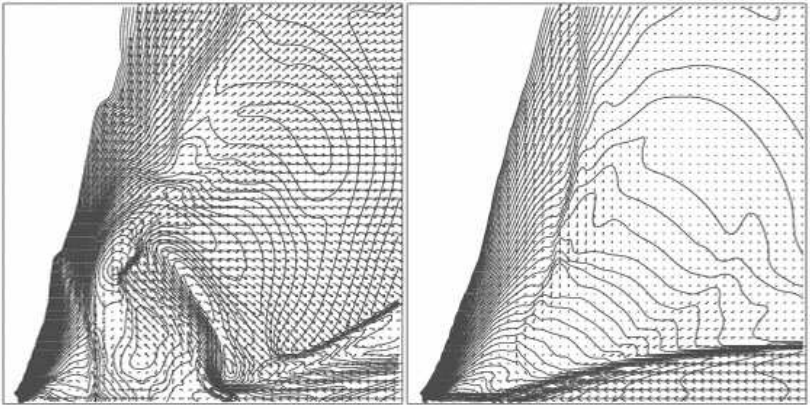}}
\vskip -0.0cm
   \caption{Comparison of the flow topologies at two phases of the shock oscillations
in simulations in which only the radial oscillation of the shocks are allowed (from RCM96). 
Here no cooling was used but the flow parameters (energy/angular momentum) have been so chosen that
it does not have a standing shock.} 
   \end{center}
\end{figure}

\begin{figure}
   \begin{center}
\vskip -0.0cm
   \mbox{\epsfxsize=0.6\textwidth\epsfysize=0.45\textwidth\epsfbox{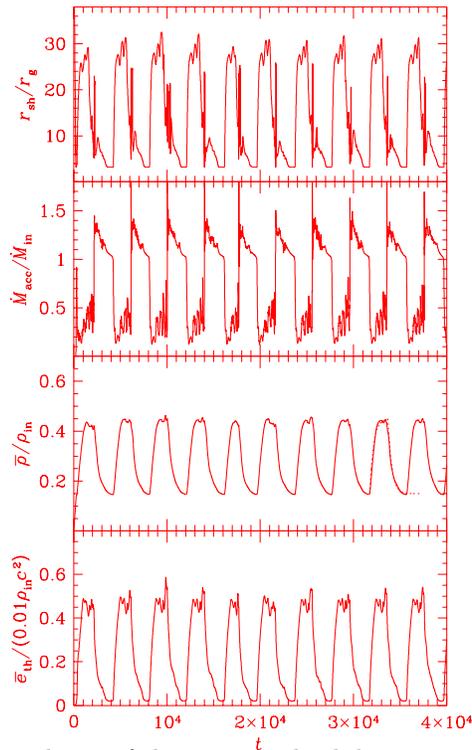}}
\vskip -1.0cm
   \caption{Time dependence of the average shock location, accretion rate, average density and average thermal energy
during sustained oscillation of an advective, sub-Keplerian disk (from RCM96). } 
   \end{center}
\end{figure}

\begin{figure}
   \begin{center}
{\vskip -3.5cm
   \mbox{\epsfxsize=0.8\textwidth\epsfysize=0.6\textwidth\epsfbox{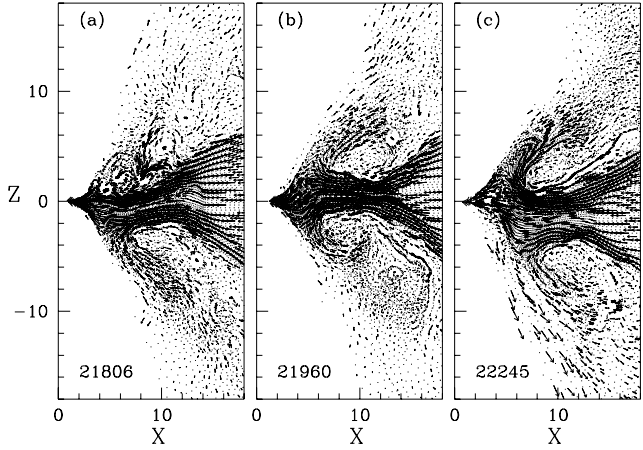}}}
\vskip -1.0cm
   \caption{Configuration of the advective flow with cooling effects at three different
phases of oscillation when both the radial and the vertical motions of the shocks are allowed (i.e.,
equatorial symmetry is also removed). The modulation of X-rays due to this oscillation causes
quasi-period oscillations in black hole candidates (from CAM04).}
   \end{center}
\end{figure}

\begin{figure}
   \begin{center}
{\vskip -1.0cm
   \mbox{\epsfxsize=0.8\textwidth\epsfysize=0.6\textwidth\epsfbox{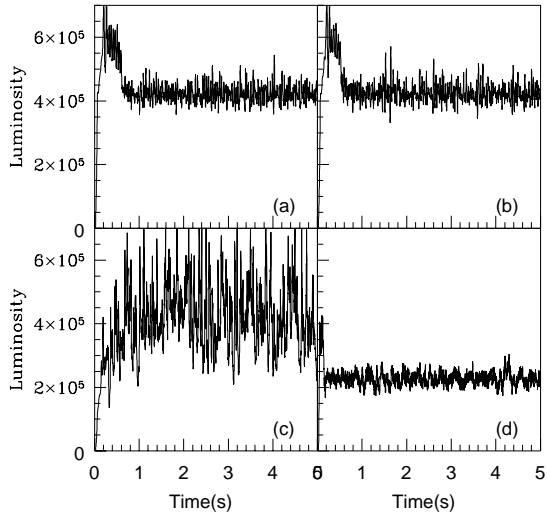}}}
\vskip -3.5cm
   \caption{Variation of emitted luminosity (in arbitrary units) with time from four numerical simulations
in increasing order of accretion rate (from CAM04). Dimensionless accretion rates (of the sub-Keplerian
flow) when the cooling is normalized to 
Compton cooling are (a) 0.05, (b) 0.08, (c) 0.39 and (d) 0.85.} 
   \end{center}
\end{figure}

\begin{figure}
   \begin{center}
\vskip 0.0cm
   \mbox{\epsfxsize=0.8\textwidth\epsfysize=0.6\textwidth\epsfbox{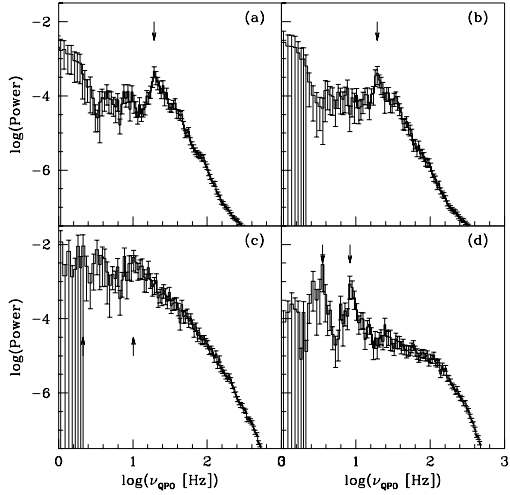}}
\vskip -3.5cm
   \caption{Variation of Power Density Spectra (PDS) for the cases presented in Fig. 5.
The QPO frequencies are (a) 19.34 Hz, (b) 19.45 Hz, (c) 10.2 Hz and 2.67 Hz and (d) 8.32 Hz and 3.58 Hz (from CAM04).}
   \end{center}
\end{figure}

\begin{figure}
   \begin{center}
\vskip -1.0cm
   \mbox{\epsfxsize=0.6\textwidth\epsfysize=0.45\textwidth\epsfbox{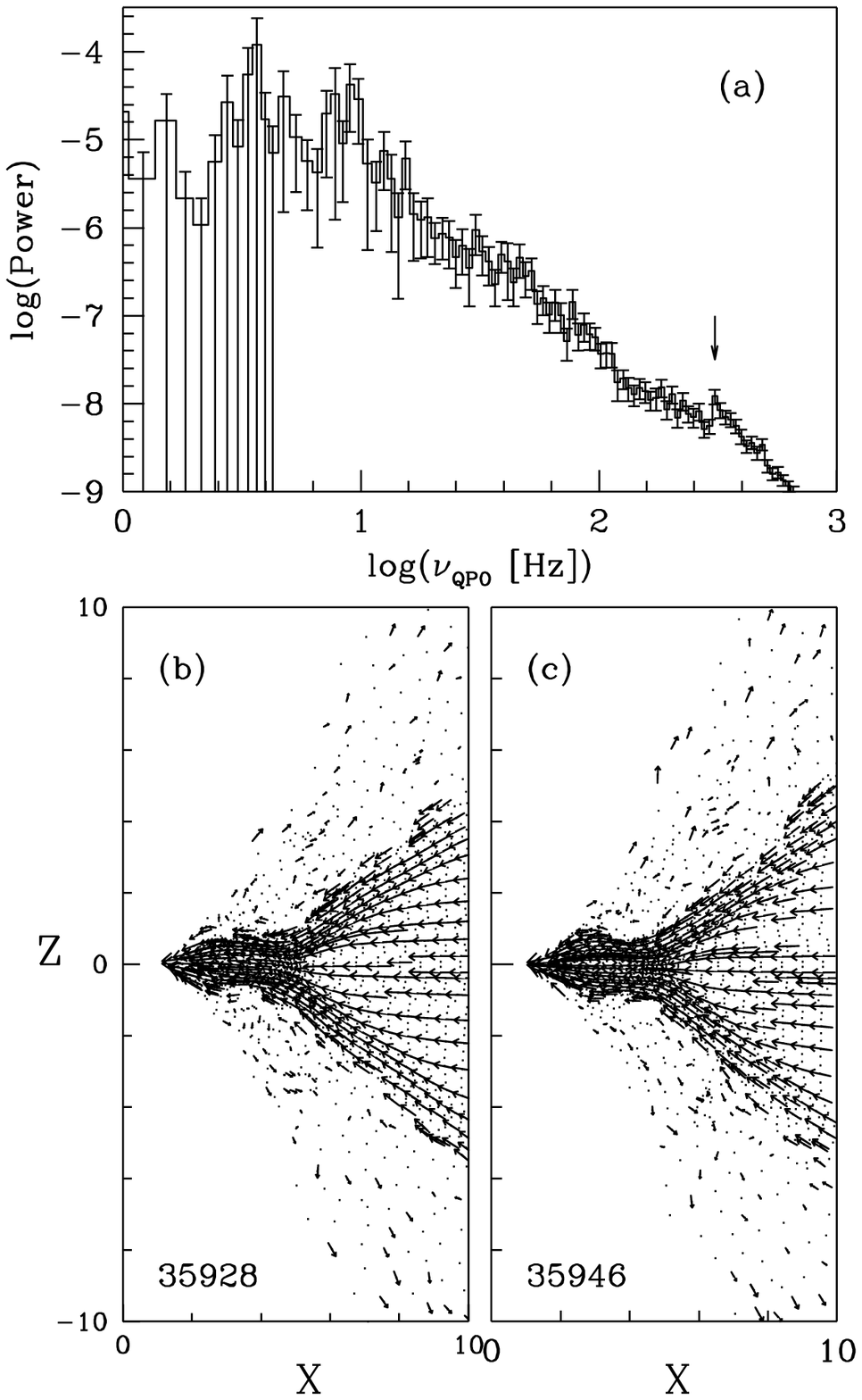}}
\vskip -0.5cm
   \caption{
The simulation results reproduces the high frequency QPO of $\nu\sim 330$Hz as seen in 
the PDS (a). The flow configurations in (b) show that this is due to the oscillation of a weak
shock just before the flow enters through the inner sonic point. }
   \end{center}
\end{figure}

Even when the explicit cooling was absent, the CENBOL was found to oscillate (Ryu, Chakrabarti \& Molteni, 
1996, hereafter RCM96) in those situations when the steady state, standing shock solution itself is absent. 
Figure 2 shows the results of a two-dimensional simulation at two different times where we see that the 
shock at its extreme locations. The top panel of the Figure 3 shows the time dependence of the shock location. 
This oscillation causes variation in the accretion rate, average density, average thermal energy of the disk. 
As pointed out in Ryu et al. (2004), the oscillation is due to not having steady solutions for the injected 
parameters. This type of oscillations should therefore be very common. 

In the advanced and the most recent work of the numerical simulation Chakrabarti, Acharyya \& Molteni, (2004)
(hereafter CAM04) showed that the power density spectra could be best reproduced if the symmetry around the
equatorial plane is relaxed and the shocks were also allowed to oscillate not only along the 
radial direction, but also along the vertical direction. Normally, in spherically symmetric space-time
these two frequencies are identical, but in axi-symmetric space-times (such as Kerr) they may be different.
The simulation in Kerr spacetime is being carried out and would be reported soon.

In Fig. 4, the configurations of the disk having both the radial and vertical oscillations of the CENBOL
are shown. In Figs. 5(a-d), we present resulting variation in emitted luminosity with time
(light curves) for four accretion rates increasing from (a) to (d). The mass of the black 
hole is chosen to be $10 M_\odot$. In Figs. 6(a-d) the power density spectra (PDS) are shown. 
It is clear that the light curves and the PDS have similar characteristics of a typical
$\chi$ class (Belloni et al. 2000). The QPO occurs at frequencies close to break-frequencies. In case (c), the 
excessive noise in the light curve resulted in the absence of any prominent QPOs.

As far as the high frequency QPOs are concerned, our understanding is that it could be due to the 
oscillation of the inner shock close to the black hole. Figure 7 (upper panel) shows the PDS of one of our
simulations which shows the presence of a high frequency QPO at around $330$Hz. The simulation
results (lower panel) indicate that there is a slight variation near the inner sonic point
at this frequency. For details, see CAM04.

\section{Concluding Remarks}

We have shown that the advective disks which include oscillating shocks reproduce both the 
low frequency and high frequency QPOs very well. The fact that QPO frequencies migrate 
with luminosity may be due to the variation of cooling time scale with accretion rate.
There are several predictions of our solution: (a) The low frequency QPOs should generally occur
during the transition to the low-hard state or during a low-hard state. (b) Since the jets are
associated with CENBOL, it is expected that when QPOs are seen, there should be 
jets/outflow activities as well. (c) Since QPOs are supposed to be due to the oscillations of the 
CENBOL which shrinks with the increase of the cooling rate, i.e., accretion rate, the 
frequency of oscillation should generally rise with accretion rate. This will be contradicted
if, say, viscosity goes down and angular momentum goes up which again increases the size of the
CENBOL. 

\acknowledgements 
This project is supported in part by ISRO project, `Quasi-Periodic Oscillations of Black Hole Candidates'
and in part by the DST project `Emitted Spectra from Two-Component Accretion Disks Around Black Holes'.

\label{lastpage}

\end{document}